\documentclass{article}
\usepackage{appendix}
\usepackage{amsmath}
\usepackage[retainorgcmds]{IEEEtrantools}
\usepackage{graphicx}

\begin{document}
\title{Bosons thermodynamics in the gravity field}
\author{Loris Ferrari* and Fabrizio Pavan\\ Department of Physics and Astronomy (DIFA) of the University of Bologna\\via Irnerio, 46 - 40126, Bologna,Italy}
\maketitle
\begin{abstract}
A uniform force like the weight has been shown to forbid Bose-Einstein condensation (BEC), due to the discretness of the Airy spectrum, resulting from the weight. We show that BEC is forbidden even if the Airy spectrum is treated as continuous, in what we call the approximate continuous limit (ACL). The absence of BEC is due to the finite difference $\epsilon_0$ between the quantum and the classical ground level. A thorough study is made, showing the way the Bose-Einstein condensate grows, at temperatures lower than the fictiuos BEC temperature $T_f$, resulting from the limit $\epsilon_0\rightarrow0$. A comparison with some experimental data shows the difference between the confinement in a rigid-wall reservoir and in a harmonic trap. 
\end{abstract}

*Corresponding author: e-mail: loris.ferrari@unibo.it

\section{Introduction}
\label{intro}
In recent years, the effects of the weight on a degenerate Boson gas passed from a merely speculative/pedagogic level, to an advanced research theme, due to the increasing progresses in ultra-cold tecniques, at temperatures as low as $1\div10^2 \mathrm{n K}$, and to the interest in using the Bose-Einstein condensate as a probe for subtle quantum effects, involving the measurement or the elimination of the gravity field \cite{Vogel,VZoest,Carraz,Frye,ZHOU,Muntinga,Meister,Bravo,McGuirk,Condon, Rudolph}. The technical feasibility of such extreme conditions imposes severe constraints (in particular, small volumes and a controlled number of particles), which make the thermodynamic limit (TL) a questionable approximation. Since TL is a \emph{necessary} condition for an ideal BEC, as shown by Yang and Lee \cite{YL1,YL2}, the application of the BEC picture to such experimental situations is open to discussion. Furthermore, in ref \cite{Soldati} it is shown that the weight itself, in a 3-dimensional gas, forbids BEC, due to the discrete nature of the Airy spectrum resulting from the weight and surviving TL. Hence, the conditions which make BEC a reliable approximation in the gravitational field deserve a careful analysis, which is the aim of the present work.  

Section \ref{ECP} is devoted to a preliminary discussion on the general aspects of BEC, including the notion of \textquoteleft effective\textquoteright$\:$ confinement volume $V_{eff}$ in the presence of external fields, and the continuous limit (CL), in which the sum over the discrete spectrum's levels is replaced by an integral of a Riemann-integrable density of states (DOS). A distinction is made between \emph{rigorous} CL (RCL), resulting from TL itself, and \emph{approximate} CL (ACL), resulting from the high-temperature approximation in which the inter-level spacings are treated as \textquoteleft infinitesimal\textquoteright$\:$increments, with respect to $\kappa T$. 

Section \ref{GRAV} deals with the effects of the weight on an otherwise free gas of Bosons in a reservoir of volume $V$. It can be seen that RCL leads to the convolution of a continuous and a discrete spectrum, which results in a staircase-wise DOS (see Fig. 1). In contrast to what is currently reported \cite{Huang,Goldstein,Lamb,deGroot,Gersch,Bagnato}, we show that BEC is impossible, even if ACL \textquoteleft smooths out\textquoteright$\:$ the steps in Fig. 1, and cures the spectrum discreteness that survives TL. This is due to the finite difference $\epsilon_0$ between the \emph{quantum} ground state and the \emph{classical} minimum energy, assumed as the energy origin. In particular, we show that the condensate population, below the fictious BEC temperature $T_f$, grows exponentially, but remains a non-extensive quantity. The deviations from an ideal BEC turn out to be definitely negligeable at room density ($10^{19}\mathrm{cm}^{-3}$), both for light (e.g. $^4\mathrm{He}$) and for heavy (e.g. $^{87}\mathrm{Rb}$) Bosons. This could not be true, however, at the very low densities achieved in ultra-cold atoms experiments on $^{87}\mathrm{Rb}$, as shown by a comparison with some experimental data from refs \cite{Condon,Rudolph}.

\section{The exact equation for the Chemical Potential}
\label{ECP}

The general equation for the chemical potential $\mu$, according to Bose-Einstein statistics reads:

\begin{equation}
\label{Eqmugen1}
1=\frac{1}{N}\overbrace{\frac{g_0}{\mathrm{e}^{\beta(\epsilon_0-\mu)}-1}}^{N_0}+\frac{1}{N}\overbrace{\int_{\epsilon_0}^\infty\frac{g(E)}{\mathrm{e}^{\beta(E-\mu)}-1}\mathrm{d}E}^{N_{exc}}\:,
\end{equation}
\\            
where $N$ is the total number of Bosons, $\beta=1/(\kappa T)$, $\epsilon_0$ is the ground state energy for the single boson, with degeneracy $g_0$, and $g(E)$ is the density of states (DOS) in energy, including Dirac $\delta$-distributions, in case of \emph{discrete} components of the spectrum. Since the population $N_0$ of the ground level can be arbitrary large, in a Bosonic gas it is convenient, if not necessary, to isolate the contribution of the ground level from that of the excited levels. 

In the TL, both $N$ and the \emph{effective} volume $V_{eff}$ occupied by the gas diverge, while the effective particle density $N/V_{eff}$ remains finite and non vanishing. For a free gas, $V_{eff}$ coincides with the volume $V$ of the reservoir, and TL reduces to the finiteness of the volume density $\rho_V=N/V<\infty$. In the presence of external fields, the confinement effects of the fields themselves make $V_{eff}$ depend on the fields' parameters and on the temperature too. In particular, the gravitational force $-mg$ in a vertical, infinitely tall cilinder, of base area $A$ yields: 

\begin{equation}
\label{Vg}
V_{eff}\approx A\frac{\kappa T}{mg}\:, 
\end{equation}
\\
in the semi-classical limit, as we shall see in what follows. Another important example is the confinement due to a 3-dimensional isotropic harmonic trap of frequency $\omega$. Since $\langle x^2\rangle_n+\langle y^2\rangle_n+\langle z^2\rangle_n=\mathcal{E}_n/(m\omega^2)$ is the mean square radius of the 3D oscillator in the $n$-th level $\mathcal{E}_n$, the confinement volume can be estimated as $(4/3)\pi\left[\langle\: \mathcal{E}\:\rangle_T/(m\omega^2)\right]^{3/2}$, $\langle\:\mathcal{E}\:\rangle_T$ being the thermal mean value of the energy. In the semi-classical limit $\omega\rightarrow0$, one has $\langle\:\mathcal{E}\:\rangle_T\rightarrow3\kappa T$, whence:

\begin{equation}
\label{Vharm}
V_{eff}\approx\frac{4\pi}{\sqrt{3}\omega^3}\left(\frac{\kappa T}{m}\right)^{3/2}\:.
\end{equation}
\\

A quantum particle confined in a finite region of space exhibits a \emph{discrete} spectrum 
$\{\epsilon_0<\epsilon_1<\cdots<\epsilon_n<\cdots\}$, with degeneracies $\{g_n\}$. The resulting DOS is a sum of $\delta$-functions. In some cases, the inter-level spacings $\epsilon_{n+1}-\epsilon_n$ tend to vanish, with diverging $V_{eff}$, so that TL itself makes it possible to replace the sum on the energy levels by an integral of a Riemann-integrable function. This is what we call a \emph{rigorous} continuous limit (RCL). An example of RCL is the harmonic trap, in which the divergence of $V_{eff}$ (eqn \eqref{Vharm}) is due to the limit $\omega\rightarrow0$ and implies the vanishing of the energy splitting $\hbar\omega$ between nearest neighbour levels. If, instead, replacing the sum with an integral follows from a high-temperature approximation $(\epsilon_{n+1}-\epsilon_n)<<\kappa T$, with fixed inter-level spacings, we speak about an \emph{approximate} continuous limit (ACL). This is actually the case of the vertical cilinder (eqn \eqref{Vg}), as we shall see in Section \ref{GRAV}. 

Having isolated the contribution of the ground energy $\epsilon_0$, it is intended that CL refers to the excited states' DOS $g(E)$. The simplest approach to CL is the semi-classical approximation, in which, given a Hamiltonian $H(\mathbf{P},\mathbf{Q})$, with ground state energy $\epsilon_0$, the number $\mathcal{N}(E)$ of eigenstates with energy less than a given value $E$, is calculated as the ratio between the volume of the classical phase-space region 

\begin{equation}
\label{Omega}
\Omega(E)=\left\{\mathbf{P},\mathbf{Q};\:\epsilon_0\le H(\mathbf{P},\mathbf{Q})<E\right\}\:,\nonumber
\end{equation}
\\
and the smallest quantum cell's volume $\mathrm{h}^D$ compatible with the Heisenberg Principle ($\mathrm{h}$ is Planck's constant). This yields (spin apart\footnote{Including a spin $s$ means multiplying by $2s+1$.}):

\begin{equation}
\label{N(E)}
\mathcal{N}(E)=\frac{1}{\mathrm{h}^D}\int_{\Omega(E)}\mathrm{d}^D\mathbf{P}\mathrm{d}^D\mathbf{Q}\:.
\end{equation}
\\
The Riemann integrable DOS resulting from eqn \eqref{N(E)} is $g(E)=\mathrm{d}\mathcal{N}(E)/\mathrm{d}E$. A special attention is to be payed to the ground state energy $\epsilon_0$, which hides a subtlety: $\epsilon_0$ is \emph{not} an energy increment, to be compared to $\kappa T$, in view of ACL, but the minimum \emph{quantum} energy, measured with respect to the \emph{classical} minimum. Therefore, $\epsilon_0$ is the basic effect of Heisenberg's principle, and cannot be cancelled by ACL.  

The ideal BEC is defined as the transition from $\mu(T)<\epsilon_0$ ($T> T_B$) to $\mu(T)=\epsilon_0$ ($T\le T_B$) at a \emph{finite} temperature  $T_B$ \footnote{The same definition of BEC applies as well, on replacing the temperature with the pressure and inverting the inequality signs, at constant $T$.}. The equation for $T_B$ follows from eqn \eqref{Eqmugen1}, for $\mu\rightarrow\epsilon_0$ from below (that means $N_0<\infty$), and $N\rightarrow\infty$:

\begin{subequations}
\label{EqTB}
\begin{equation}
\label{EqTB1}
1=\int_{\epsilon_0}^\infty\frac{g(E)}{N(\mathrm{e}^{\beta_B (E-\epsilon_0)}-1)}\mathrm{d}E\quad;\quad \beta_B=1/(\kappa T_B)\:.
\end{equation}
\\
For $T_B$ to be non vanishing and finite, the integral in $E$ must be bounded for any value of $\beta_B>0$ strictly. On expanding the exponential about $\epsilon_0$, it is easy to see that this yields:

\begin{equation}
\label{CondBEC}
\mathrm{lim}_{\xi\rightarrow0}\int_{\xi+\epsilon_0}^{\delta}\frac{g(E)}{N(E-\epsilon_0)}\mathrm{d}E<\infty\quad\text{for each }\epsilon_0<\delta<\infty\:.
\end{equation}
\end{subequations}
\\
The condition \eqref{CondBEC} is necessary (and sufficient) for a rigorous BEC, whose physical meaning is the transition of the ground level population $N_0$ (eqn \eqref{Eqmugen1}) from being a \emph{finite} quantity, for $\mu(T)<\epsilon_0$ (strictly), to be \emph{extensive} ($\propto N$), for $\mu(T)=\epsilon_0$.

\section{Adding a gravity field to the free gas}
\label{GRAV}
A uniform gravity force $-mg$ acting along the z-axis of an otherwise free gas yields the single-particle Hamiltonian:

\begin{align}
H&=\frac{p_x^2+p^2_y}{2m}+\overbrace{\frac{p^2_z}{2m}+mgz}^{H_z},\quad \text{for } (x,\:y)\in\mathcal{A},\:\:0\le z \le L\label{H}\\
\nonumber\\
&=\infty\quad\text{otherwise}\:,\nonumber
\end{align}
\\
where $\mathcal{A}$ is a bounded 2-dimensional region of area $A$ (the base), and $L$ is the height of the reservoir. The spectrum of the pure kinetic energy in the $x,y$-plane becomes \emph{rigorously continuous} (RCL) in the limit of divergingly large area $A$ of the reservoir's base $\mathcal{A}$, with excited states' DOS (see, for instance, ref. \cite{Soldati}):

\begin{equation}
g_{A}(\epsilon)=
\begin{cases}\underbrace{2m/(\pi\hbar)}_{g_{2d}}A\quad&\text{for }\epsilon>0\nonumber\\
\label{GA}\\
0\quad\quad\quad\quad\quad&\text{for }\epsilon<0\:,\nonumber
\end{cases}
\end{equation}
\\
and ground level $\epsilon_m=0$ with degeneracy 1\footnote{The ground level for the kinetic energy in a 2D box of area $A$ is $\epsilon_m=\hbar^2\pi^2/(2mA)$ and vanishes in the TL.}. In the present case TL yields $A,\:N\rightarrow\infty$, with a finite surface density $\rho_A=N/A$. In contrast, the Airy-Schr\"{o}dinger equation, that solves the quantum eigenvalue problem for $H_z$, yields an intrinsically \emph{discrete} spectrum

\begin{align}
\label{epsilonn}
&\epsilon_n=\overbrace{\left(m\mathrm{h}^2g^2\right)^{1/3}}^{E_0}\left(\frac{9}{32}\right)^{1/3}\left(n+\frac{3}{4}\right)^{2/3}\times\Bigg[1+\nonumber\\
\nonumber\\
&+\frac{20}{48(3\pi^2)(n+3/4)^2}-\frac{20}{36(3\pi^4)(n+3/4)^4}+\circ(1/n^6)\Bigg]\quad n=0,\:1,\:\cdots\:,
\end{align}
\\
resulting from the boundary conditions $\psi_n(0)=\psi_n(\infty)=0$, on the eigenfunctions $\psi_n(z)$ \cite{Wheeler}. Since there is no dependence of $\epsilon_n$ on $A$, the discreteness of the spectrum eqn \eqref{epsilonn} cannot be removed by the limit $A\rightarrow\infty$. For the total spectrum to become rigorously continuous, in fact, one should add the condition $g\rightarrow0$, which we ignore for the moment. In conclusion, the exact DOS corresponding to $H_z$ is a series of Dirac $\delta$-distributions:

\begin{equation}
\label{Gz}
g_z(\epsilon')=\sum_{n=0}^\infty\delta\left(\epsilon'-\epsilon_n\right)\:,
\end{equation}
\\  
since the degeneracy of any level is $g_n=1$ for each $n$. From eqns \eqref{GA} and \eqref{Gz}, one gets the exact DOS for the excited states in the total energy $E=\epsilon+\epsilon'>\epsilon_0$:

\begin{align}
\label{Gex}
g_{ex}(E)&=\int_0^\infty\mathrm{d}\epsilon\int_0^\infty\mathrm{d}\epsilon' g_A(\epsilon)g_z(\epsilon')\delta\left(E-\epsilon-\epsilon'\right)=\nonumber\\
\nonumber\\
&=g_{2d}A\sum_{n=0}^\infty\int_0^\infty\delta\left(E-(\epsilon+\epsilon_n)\right)\mathrm{d}\epsilon\quad(E>\epsilon_0)
\end{align}
\\
(see Fig. 1). Equations \eqref{epsilonn}, \eqref{Gz} and \eqref{Gex} refer to an infinitely high reservoir ($L\rightarrow\infty$), in which the gas is confined by the gravitational force, even in the absence of an upper cover. In a reservoir of finite height $L$, the different boundary condition $\psi_n(L)=0$ would change the eigenvalues' expression \eqref{epsilonn}. For macroscopic values of the height such modifications can be accounted for by setting an upper limiting value $M(L)$ to the \emph{number of eigenvalues} $\epsilon_n$ involved in the calculations, i.e. to the index $n$. On applying the Virial Theorem to the Airy-Schr\"{o}dinger equation, it is seen that the mean kinetic energy equals the mean potential energy \cite{Ipecoglu}. So, one can write, for the particle's mean height $\langle\:z\:\rangle_n$ in the $n$-th eigenstate:

\begin{figure}[htbp]
\begin{center}
\includegraphics[width=5in]{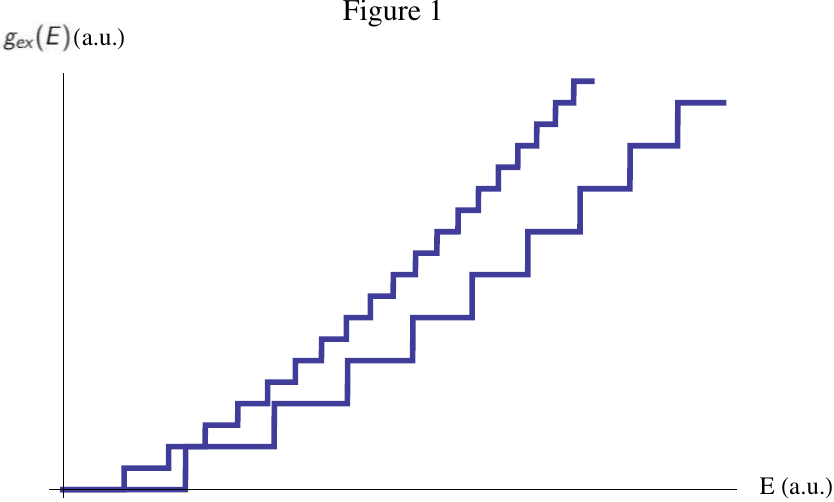}
\caption{\textbf{Riemann-integrable part of the single particle DOS under the action of a uniform force (arbitrary units)}. The steps are due to the discrete spectrum of $H_z$ and become smaller and smaller with decreasing $E_0$ (eqn \eqref{epsilonn}). The limit for $E_0\rightarrow0$ is a continuous DOS proportional to $E^{3/2}$.}
\label{default}
\end{center}
\end{figure}

\begin{equation}
mg\langle\:z\:\rangle_n= \frac{\epsilon_n}{2}\:,\nonumber
\end{equation}
\\
and apply the condition $\langle\:z\:\rangle_n\le L$ to determine $M(L)$:

\begin{equation}
\epsilon_{M(L)}\simeq2mgL\quad \Rightarrow\quad M(L)\simeq \mathrm{Int}\left(\frac{2mgL}{E_0}\right)^{3/2}\quad \text{for }M(L)>>1\:,\nonumber
\end{equation}
\\
according to eqn \eqref{epsilonn}. The effective volume, in the present case, is: 

\begin{equation}
V_{eff}=A\langle\:z\:\rangle_T=A\sum_{n=0}^{M(L)}\langle\:z\:\rangle_n\mathrm\:{e}^{-\beta\epsilon_n}g_n/\sum_{n=0}^{M(L)}\mathrm\:{e}^{-\beta\epsilon_n}g_n\:,\nonumber
\end{equation}
\\
where $g_n=1$ and the thermal mean value $\langle\:z\:\rangle_T$ represents the average thickness of the layer of particles lying on the reservoir's bottom. For $L\rightarrow\infty$ and in the semi-classical limit, one readily gets $\langle\:z\:\rangle_T\rightarrow\kappa T/(mg)$, which leads to eqn \eqref{Vg} for the effective volume. 

Given the number $N$ of spinless bosons in the reservoir, the equation for the chemical potential $\mu$ reads, from eqn \eqref{Gex}:
\begin{subequations}
\begin{align} 
&1=\frac{N_0}{N}+\frac{1}{N}\int_0^\infty g_{ex}(E) \frac{\mathrm{d}E}{\mathrm{e}^{\beta(E-\mu)}-1}\Rightarrow\label{Eqmu1}\\
\nonumber\\
\:\:\Rightarrow\:\: &1=\frac{N_0}{N}+g_{2d}\frac{A}{N}\sum_{n=0}^{M(L)}\int_0^\infty\frac{\mathrm{d}\epsilon}{\mathrm{e}^{\beta(\epsilon+\epsilon_n-\mu)}-1}\:.\label{Eqmu2}
\end{align}
\end{subequations}
\\
On defining:

\begin{subequations}
\label{Deltan,EqDeltamu}
\begin{equation}
\label{Deltan}
\Delta\mu:=\epsilon_0-\mu\ge0\quad,\quad\Delta_{n}:=\epsilon_{n+1}-\epsilon_0>0,\quad \text{for }n=0,\:1\:,\cdots\:,
\end{equation}
\\
equation \eqref{Eqmu2} becomes:

\begin{equation}
\label{EqDeltamu}
1= \frac{N_0}{N}+g_{2d}\frac{A}{N}\left[\int_0^\infty\frac{\mathrm{d}\epsilon}{\mathrm{e}^{\beta(\epsilon+\Delta\mu)}-1}+\sum_{n=0}^{M(L)}\int_0^\infty\frac{\mathrm{d}\epsilon}{\mathrm{e}^{\beta(\epsilon+\Delta_n+\Delta\mu)}-1}\right]\:.
\end{equation}
\end{subequations}
\\
Note that $\Delta\mu$ is \emph{non negative}, since $N_0$ (eqn \eqref{Eqmugen1}) cannot be negative. For BEC to occur, $\Delta\mu$ must vanish at a \emph{finite} temperature $T_B$, which yields $\beta_B<\infty$. In eqn \eqref{EqDeltamu} this is impossible (whatever the value of $N_0$), because $A/N$ is a non vanishing quantity (due to TL), and the first integral diverges for any finite value of $\beta$, if $\Delta\mu=0$. Hence, the gravity field forbids BEC, because the resulting DOS \eqref{Gex} is \emph{non vanishing}, down to the ground level $\epsilon_0$. A similar argument is used in ref. \cite{Soldati} to deny BEC in a uniform force field. 

\subsection{The approximate continuous limit (ACL) and the gravity field}
\label{GRAV2}

Despite the discreteness of the spectrum eqn \eqref{epsilonn} survives TL, an approximate continuous limit (ACL) can be performed, as mentioned in Section \ref{ECP}, if the differences between consecutive levels can be treated as \textquoteleft infinitesimal\textquoteright$\:$ increments. This means assuming that such differences are small compared to the thermal energy scale $\kappa T$, which yields: 

\begin{equation}
\label{acl}
\beta(\epsilon_{n+1}-\epsilon_n)<<1\quad \text{for each } n=0,\:1,\:\cdots\:.
\end{equation}
\\
According to eqn \eqref{epsilonn}, $\Delta_1=E_0/\kappa\times 0.398$ is the largest difference between consecutive eigenvalues, hence the condition \eqref{acl} reads, in the gravity field ($g=9.81\times10^2\:\mathrm{cm}/\mathrm{s}^2$):

\begin{equation}
\label{ACL2}
T>>\Delta_1/\kappa=
\begin{cases}
1.89\times 10^{-8}\:\mathrm{K}&\text{ for }^4\mathrm{He}\:(m=6.65\times 10^{-24}\mathrm{g})\\
\\
5.28\times 10^{-8}\:\mathrm{K}&\text{ for }^{87}\mathrm{Rb}\:(m=1.454\times 10^{-22}\mathrm{g})\:,
\end{cases}
\end{equation}
\\
for a light ($^4\mathrm{He}$) and a heavy ($^{87}\mathrm{Rb}$) Bosonic atom, currently used in experiments.\newline

In the case eqn \eqref{H}, equation \eqref{N(E)} yields the following DOS for the excited states (see Appendix \ref{Appendix A}):

\begin{align}
g_{acl}(E,\:L,\:g)&=\frac{4\pi(2m)^{3/2}A}{3mg\mathrm{h}^3}E^{3/2}\times\nonumber\\
\nonumber\\
\times
&\begin{cases}
1-\left(1-\frac{mgL}{E}\right)^{3/2}\quad&\text{for }E\ge mgL\\
\\
1\quad&\text{for }\epsilon_0< E\le mgL\label{G(E,L)}\\
\\
0\quad&\text{for }E< \epsilon_0\:,
\end{cases}
\end{align}
\\
where the ground state energy $\epsilon_0$ is explicitly included. Equation \eqref{G(E,L)} shows the dependence of DOS on the \emph{shape} of the reservoir, given that the height $L$ and the area $A$ can be modified independently. In particular, the limit $g\rightarrow0$ of eqn \eqref{G(E,L)} is an example of RCL and describes a weightless gas, whose confinement along the $z$-axis is completely determined by the upper cover. It is left to the reader to show that for $g\rightarrow0$ equation \eqref{G(E,L)} yields the DOS of the \emph{perfect gas} in 3d:

\begin{align}
g_{acl}(E,\:L,\:0)&=\frac{2\pi(2m)^{3/2}V}{\mathrm{h}^3}E^{1/2}\quad\text{for }E> 0\nonumber\\
\\\label{G0}
&=0\:\quad\quad\quad\quad\quad\quad\quad\quad\text{for } E<0\nonumber
\end{align}
\\
(note that $\epsilon_0\rightarrow0$ for $g\rightarrow0$ and $V=AL$). The study of BEC in the perfect gas is a current issue in most textbooks of statistical thermodynamics, which yields: 

\begin{subequations}
\begin{equation}
\beta\Delta\mu=\beta|\mu|=
\begin{cases}
C_{1/2}\left(\frac{T-T_B}{T_B}\right)^2+\cdots&\quad\text{for }T\ge T_B \\
\label{mu_0}\\
0&\quad\text{for }T\le T_B\:,
\end{cases}
\end{equation}
\\
and $T$ close to the condensation temperature:

\begin{equation}
\label{T0}
T_B=\frac{\mathrm{h}^2}{2\kappa\pi m}\left[\frac{\rho_V}{\zeta(3/2)}\right]^{2/3}\:,
\end{equation}
\end{subequations}
\\   
$\rho_V=N/V$ being the volume density and $\zeta(\cdot)$ the Riemann function. For the interested reader, the numerical factor $C_{1/2}=1.222$ can be calculated from the general formulas reported in the appendix of ref. \cite{Me}. 

The presence of the weight prevents the vanishing of $\epsilon_0$, whence $g_{acl}(\epsilon_0,\:L,\:g)>0$ (eqn \eqref{G(E,L)}) and:

\begin{equation}
\mathrm{lim}_{\xi\rightarrow0}\int_{\epsilon_0+\xi}^\delta\frac{g_{acl}(E,\:L,\:g)}{N(E-\epsilon_0)}\mathrm{d}E=\infty\quad \text{for each } L\:,
\end{equation}
\\ 
which violates condition \eqref{CondBEC} and forbids BEC for any value of the reservoir's height, \emph{even if} ACL is assumed. So, it is not the discreteness of the spectrum \eqref{epsilonn} that cancels BEC, but the non vanishing behavior of the DOS at the ground level $\epsilon_0$ \footnote{This result might sound counter-intuitive, at a first sight, since in a semi-classical approximation, with the minimum possible quantum content, one expects that the weight should \emph{favor} BEC, in that the weight itself \textquoteleft pushes\textquoteright$\:$ the gas down to the ground state. The crucial point is the unavoidable quantum effect hidden in $\epsilon_0$, i.e., the Heisenberg principle.}. The incorrect assumption (see below) that ACL makes $\epsilon_0$ a negligeable quantity, transforms \eqref{G(E,L)} in a \textquoteleft fictious\textquoteright$\:$DOS:

\begin{align}
g_{f}(E,\:L)&=\frac{4\pi(2m)^{3/2}A}{3mg\mathrm{h}^3}E^{3/2}\times\nonumber\\
\nonumber\\
\times
&\begin{cases}
1-\left(1-\frac{mgL}{E}\right)^{3/2}\quad&\text{for }E\ge mgL\\
\\
1\quad&\text{for }0\le E< mgL\label{Gf}\\
\\
0\quad&\text{for }E<0\:,
\end{cases}
\end{align}
\\ 
which satisfies the BEC condition \eqref{CondBEC}. This makes BEC \textquoteleft resuscitate\textquoteright, as reported in many textbooks and articles \cite{Huang,Goldstein,Lamb,deGroot,Gersch,Bagnato}, which use ACL in the presence of a uniform force field, ignoring the intrinsic discreteness of the $H_z$ spectrum and the existence of a ground level $\epsilon_0>0$, above the minimum classical energy. In order to recover an exact BEC, one should take $L\rightarrow\infty$, $g\rightarrow0$, $A\rightarrow\infty$ in eqn \eqref{G(E,L)}. In addition, TL should read $N\propto A/g$. In this case, the resulting critical behavior of $\beta\Delta\mu$, close to the fictious condensation temperature $T_f$, turns out to be

\begin{subequations}
\begin{equation}
\beta\Delta\mu=\beta|\mu|=
\begin{cases}
C_{3/2}\left(\frac{T-T_f}{T_f}\right)+\cdots&\quad\text{for }T\ge T_f \\
\label{mu_f}\\
0&\quad\text{for }T\le T_f\:,
\end{cases}
\end{equation}
\\
with

\begin{equation}
\label{Tf}
T_f=\frac{1}{\kappa}\left[\frac{g^2\:\mathrm{h}^6\:\rho_A^2}{8\pi^3m\zeta^2(5/2)}\right]^{1/5}\:,
\end{equation}
\\
$\rho_A=N/A$ being the surface density. According to the current literature, $T_f$ in eqn \eqref{Tf} is calculated for an infinitely tall reservoir, as the solution of the equation

\begin{equation}
\label{EqTf}
1=\frac{4\pi(2m)^{3/2}}{3mg\mathrm{h}^3}\frac{A}{N}\int_0^\infty\mathrm{d}E\frac{E^{3/2}}{\mathrm{e}^{\beta_fE}-1}\:,
\end{equation}
\end{subequations}
\\
for $\beta_f=1/(\kappa T_f)$. The numerical factor $C_{3/2}=1.284$ follows in turn from ref. \cite{Me}. Here we stress the relevant aspect of eqns \eqref{mu_0} and \eqref{mu_f}, i.e. the \emph{quadratic} and \emph{linear} approach to zero of $\beta\Delta\mu$ in the two cases. Those are special examples of a general rule expressed in ref. \cite{Me} (see eqn (2) and Fig. 1 therein\footnote{In the figure indicated there is a misprint: the symbol $\lambda$ is to be replaced by $\gamma$}). In view of what follows, it should not escape from attention that all the preceding formulas, concerning the fictious BEC, implicitly assume that ACL does apply down to $T_f$. According to eqn \eqref{ACL2}, \eqref{epsilonn} and \eqref{Tf}, this leads to a condition on the surface density $\rho_A=A/N$:

\begin{equation}
\label{CondACL}
\kappa T_f>>\Delta_1\Rightarrow0.735\left[\frac{m^2g}{\mathrm{h}^2}\left(\frac{A}{N}\right)^{3/2}\right]^{4/15}<<1\:.
\end{equation}
\\ 
 
\subsection{Using the fictious DOS appropriately}
\label{EPC3}

The results obtained in what precedes show that taking $\epsilon_0=0$ as a consequence of the ACL condition $\beta\Delta_1<<1$, leads one to neglect a \emph{diverging large} piece of integral, despite $\beta\epsilon_0<<1$ too, since $\epsilon_0\approx\Delta_1$ (eqn \ref{epsilonn}). The conclusion is that it is impossible to realize the BEC condition $\Delta\mu=0$ at any finite temperature, in the presence of a uniform force, no matter if ACL is assumed or not. However, the fictious DOS $g_{f}(E)$ can be conveniently used to locate the parameters' region in which the approximation $\epsilon_0=0$ definitely fails. Let us deal with the limit $L\rightarrow\infty$, in which the confinement of the gas is totally due to the weight. From eqn \eqref{G(E,L)} it follows that: 

\begin{subequations}
\begin{align}
1&=\frac{N_0}{N}+\int_{0}^\infty\frac{g_{acl}(E)}{\mathrm{e}^{\beta(E-\Delta\mu)}-1}\mathrm{d}E\quad\Rightarrow\nonumber\\
\nonumber\\
\Rightarrow1&=\frac{N_0}{N}+\frac{4\pi(8m)^{1/2}}{3g \mathrm{h}^3}\frac{A}{N}\int_0^{\infty}\frac{(E+\epsilon_0)^{3/2}}{\mathrm{e}^{\beta(E+\Delta\mu)}-1}\mathrm{d}E\:.\label{total}
\end{align}
\\
Since it has been shown that $\Delta\mu>0$ strictly, for any $T>0$, the population $N_0$ is finite and the term $N_0/N$ vanishes in the TL, whence:

\begin{align}
1&=\frac{4\pi(8m)^{1/2}(\kappa T)^{5/2}}{3g \mathrm{h}^3}\frac{A}{N}\Bigg[\int_0^{\infty}\frac{x^{3/2}}{\mathrm{e}^{x+\beta\Delta\mu}-1}\mathrm{d}x\:+\label{unperturbed}\\
\nonumber\\
&+\underbrace{\sum_{n=1}^\infty(\beta\epsilon_0)^n\frac{(3/2)(3/2-1)\cdots(3/2-n+1)}{n!}\int_0^{\infty}\frac{x^{3/2-n}}{\mathrm{e}^{x+\beta\Delta\mu}-1}\mathrm{d}x}_{\text{remainder}}\Bigg]\:\label{remainder},
\end{align}
\end{subequations}
\\
where the passage from eqn \eqref{total} to eqns \eqref{unperturbed}, \eqref{remainder} follows from the McLaurin expansion of $(E+\epsilon_0)^{3/2}$ in series of $\epsilon_0$. For $\epsilon_0$ to be treated as a perturbation, the remainder eqn \eqref{remainder} must be small. However, it is immediately seen that for $n\ge2$ all the integrals in the series \emph{diverge} in the limit $\beta\Delta\mu\rightarrow0$. Hence the condition for the remainder to be small, even in the limit $\beta\Delta\mu<<1$, reads: 

\begin{equation}
\beta\epsilon_0<<\:\beta\Delta\mu<<1\label{betaepsilon1<<}\:.
\end{equation}
\\
In this case, $\epsilon_0$ can be treated as a perturbation, whose effect on the unperturbed equation \eqref{unperturbed} does not change the order of magnitude of the solution for $\beta\Delta\mu$. Hence, in accounting for condition \eqref{betaepsilon1<<}, one can compare $\beta\epsilon_0$ with the fictious $\beta\Delta\mu$, resulting from eqn \eqref{mu_f}, which yields (numerical factors of order unity apart):

\begin{equation}
\frac{\epsilon_0}{\kappa T}<<\frac{T-T_f}{T_f}<<1\nonumber\:.
\end{equation}
\\
Since the second inequality implies that $T\approx T_f$, the first inequality yields:

\begin{subequations}
\begin{equation}
\label{epsilon1<<}
T-T_f>>\epsilon_0/\kappa\:,
\end{equation}
\\
and:

\begin{align}
\frac{\epsilon_0}{\kappa T_f}<<1\quad\Rightarrow&\quad1>>1.031\left[\frac{m^2g}{\mathrm{h}^2}\left(\frac{A}{N}\right)^{3/2}\right]^{4/15}\Rightarrow\label{r}\\
\nonumber\\
\Rightarrow&\quad\rho_A:=\frac{N}{A}>>0.98\left[\frac{m^2g}{\mathrm{h}^2}\right]^{2/3}\label{rho>>}\:,
\end{align}
\end{subequations}
\\
according to eqns \eqref{ACL2}, \eqref{epsilonn}, \eqref{Tf}. In conclusion, the region of the parameters space $\left(\rho_A,\:T\right)$ in which the fictious DOS can be safely used even for \emph{small} values of $\beta\Delta\mu$ is determined by two lower limits: one in $\rho_A$ (eqn \eqref{rho>>}), and one in $T-T_f$ (eqn \eqref{epsilon1<<}). For a gas of $^4\mathrm{He}$ and one of $^{87}\mathrm{Rb}$ the lower limits eqn \eqref{rho>>} are about $10^5\:\mathrm{cm}^{-2}$, and $3\times10^8\:\mathrm{cm}^{-2}$, respectively. Both values are extremely small, with respect  to what one expects from reservoir's sizes and gas densities in standard laboratory conditions. For instance, in a cubic box of side $L\approx 10\:\:\mathrm{cm}$, $\rho_A$ is about $10^{20}\:\mathrm{cm}^{-2}$ for a typical perfect gas density at room temperature and pressure ($\rho_V\approx10^{19}\:\mathrm{cm}^{-3}$). This means that the fictious BEC is a quite good approximation even at very low temperatures, except for critically low densities of bosons. 

A numerical factor apart, the lower limiting value \eqref{rho>>} for the surface density $\rho_A$ has the same order of magnitude as the one obtained from the ACL condition \eqref{CondACL}, which is not surprising, since $\epsilon_0\approx\Delta_1$. However, the ACL condition $T_f>>\Delta_1/\kappa$ does not include the condition eqn \eqref{epsilon1<<} for the fictious BEC to be a reliable approximation. Actually, $T-T_f$ can be comparable or even much smaller than $\epsilon_0/\kappa$, even if $T_f>>\Delta_1/\kappa\approx\epsilon_0/\kappa$. This should conclusively cancel the incorrect feeling that ACL includes the approximation $\epsilon_0\rightarrow0$. 

Studying what happens in the region $\Delta\mu\le\epsilon_0$, makes it possible to understand the way the true behavior of $\Delta\mu$ differs from an ideal BEC. If $\Delta\mu\le\epsilon_0$, the remainder  eqn \eqref{remainder} does not vanish, for $\Delta\mu\rightarrow0$. So, we split the r.h.s. member of eqn \eqref{total} into two parts: one diverging and one converging, for $\beta\Delta\mu\rightarrow0$. It is clear that the former yields the overhelming contribution to the solution, just when $\Delta\mu\le\epsilon_0$. As shown in Appendix \ref{Appendix B} (eqn \eqref{theta}), this leads to the following equation:

\begin{subequations}
\begin{align}
-\ln (\beta\Delta\mu)&=\mathcal{C}\left[\frac{1-(T/T_f)^{5/2}}{T/T_f}\right]+\cdots\:\label{thetatext}\\
\nonumber\\
\mathcal{C}&=\Gamma(5/2)\zeta(5/2)\left(\frac{\kappa T_f}{\epsilon_0}\right)^{3/2}=2.258\left[\left(\frac{N}{A}\right)^{3/2}\frac{\mathrm{h}^2}{m^2g}\right]^{2/5}\:,\label{C}
\end{align}
\end{subequations}
\\

\begin{figure}[htbp]
\begin{center}
\includegraphics[width=5in]{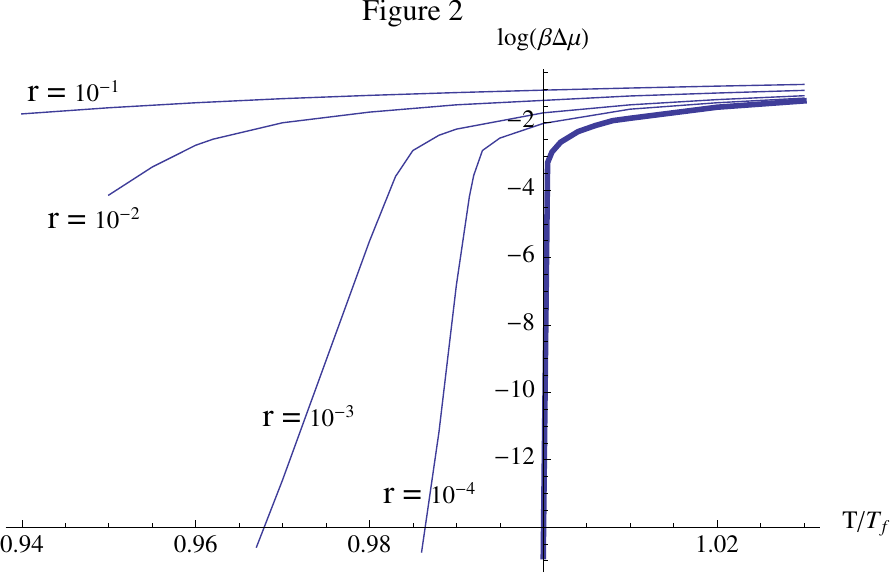}
\caption{\textbf{Absence of BEC despite ACL due to the ground energy $\epsilon_0$}. $\mathrm{log}(\beta\Delta\mu)$ is plotted against the ratio $T/T_f$ between temperature and the fictious BEC temperature $T_f$, for different orders of magnitude of the ratio $r=\epsilon_0/(\kappa T_f)$. Note the linear negative slope below $T_f$. The thick line shows the ideal BEC.}
\label{default}
\end{center}
\end{figure}

where $\cdots$ stand for terms negligible in the limit $\beta\Delta\mu<<1$ and $\beta\epsilon_0<<1$. Since equation \eqref{thetatext} has been obtained in the limit $\beta\Delta\mu<<1$, it follows that the formula itself must be used for $T<T_f$ only, i.e. in the region where $\Delta\mu=0$, according to the fictious BEC. The correct use of ACL, instead, shows an exponential decay to zero of $\beta\Delta\mu$, for $T\rightarrow0$, with exponential rate $\mathcal{C}$ (eqn \eqref{C}). The difference between the two cases is well displayed in Fig. 2, which shows $\log(\beta\Delta\mu)$ as a function of $T/T_f$. While the fictious BEC predicts a negative divergence for $T\rightarrow T_f$ from above, the correct ACL yields a linear decrease of the logarithm below $T_f$, with a coefficient $\mathcal{C}$ decreasing with the ratio $r:=\epsilon_0/(\kappa T_f)$ and diverging for $r\rightarrow0$, so that the population of the ground level increases from negligible to exponentially large values

\begin{equation}
\label{N0}
N_0=\frac{1}{\mathrm{e}^{\beta\Delta\mu}-1}=\exp\left(\mathcal{C}\left[\frac{1-(T/T_f)^{5/2}}{T/T_f}\right]\right)\quad;\quad T<T_f\:,
\end{equation}
\\
just below the fictious temperature $T_f$. Note, however, that $N_0$ is \emph{not} an extensive quantity, as it would be in an ideal BEC. This is the non trivial way by which BEC is recovered in the limit $\epsilon_0\rightarrow0$.  

It is interesting to notice from eqn \eqref{C}, that the condition $\mathcal{C}>>1$ is essentially the same (numerical factors apart) as eqn \eqref{rho>>}. Hence $\mathcal{C}$ is actually very large, apart from critically low values of the surface density $\rho_A=N/A$. In standard laboratory conditions, the transition is practically indistinguishable from an ideal BEC. For the deviations from an ideal BEC to be significant, Figure 2 shows that the ratio $r$ (eqn \eqref{r}) should be of order unity, which yields $\rho_A\approx10^8\mathrm{cm}^{-2}$ for $^{87}\mathrm{Rb}$. A question that deserves some attention is the \emph{volume} density $\rho_{eff}=N/V_{eff}$ resulting from such a small surface density, in the experimental conditions typical of the gravimetric experiments. The question of interest is actually studying the behavior of the heavy gas in a vertical cilinder, in the most extreme conditions of temperature and density, achievable in a real experiment. For instance, in ref \cite{Condon} a gas of about $N=10^5$ atoms of $^{87}\mathrm{Rb}$, is confined in a non isotropic dipole trap, realizing a harmonic potential well. It is well known \cite{Davis,Bradley, Romero} that a 3-dimensional harmonic potential (here assumed isotropic, for simplicity) of frequency $\omega$ yields a finite BEC temperature

\begin{equation}
\label{T0trap}
T_B=\frac{\hbar\omega}{\kappa}\left(\frac{2N}{\zeta(3)}\right)^{1/3}\:,
\end{equation} 
\\
which attains a value $T_B\approx 200\mathrm{nK}$, in the experimental setup of ref \cite{Condon}. From eqn \eqref{T0trap}, one gets $N=(\kappa T_B/\hbar\omega)^3\zeta(3)/2$, whence equation \eqref{Vharm} yields

\begin{equation}
\rho_{eff}^{ex}\approx\frac{\zeta(3)}{8\pi\sqrt{3}}\left(\frac{m\kappa T_B}{\hbar^2}\right)^3\approx10^{10}\mathrm{cm}^{-3}\:,
\end{equation}
\\
as a rough estimate of the experimental effective density in the harmonic trap at $T=T_B$. At the same temperature, the effective density corresponding to $\rho_A\approx10^{8}\mathrm{cm}^{-2}$ in the vertical cilinder is (eqn \eqref{Vg}):

\begin{equation}
\rho_{eff}^{c}\approx\rho_A\frac{mg}{\kappa T_B}\approx5\times10^{11}\mathrm{cm}^{-3}\:.
\end{equation} 
\\
Data extracted from the text of ref \cite{Rudolph} indicate that the effective density, in the early stages of the experiment described therein, attains values of $10^{12}\mathrm{cm}^{-3}$, comparable with $\rho_{eff}^{c}$. For densities much larger than $\rho_{eff}^c$, the increase of the condensate population in the vertical cilinder is sharp, and BEC is a good approximation. Since, instead, $\rho_{eff}^{ex}$ is much smaller than, or comparable to $\rho_{eff}^{c}$ \cite{Condon,Rudolph}, we can argue that if the harmonic trap were replaced by a vertical cilinder, in the gravimetric experiments, not only BEC would be a quite unsuitable approximation, but getting a significant increase of the condensate population would require a much lower temperature than for the harmonic trap.

\section{Conclusions}

Recent experimental techniques made it possible to observe the effects of the transition from a scarcely to a densely populated ground level (and \emph{vice versa}), for a controlled number $N$ of particles, in a microscopic effective volume $V_{eff}$. This is generally indicated as a \textquoteleft BEC\textquoteright, despite such technical constraints raise the question to what extent the notion of ideal BEC is appropriate to describe the concrete experimental situation. In particular, experiments involving the measurements or the elimination of the gravity effects have been performed, using the Bose-Einstein condensate as a probe \cite{Vogel, VZoest,Carraz,Frye,ZHOU,Muntinga,Meister,Bravo,McGuirk,Condon, Rudolph}. However, the presence of the weight itself has been shown to forbid BEC, even in cases in which TL does apply \cite{Soldati}. In the present work, the absence of BEC in heavy Boson gases has been studied in more detail. In Section \ref{ECP}, we have introduced the notion of effective confinement volume $V_{eff}$ in the presence of external fields, which is relevant for the definition of thermodynamic limit (TL) and continuous limit (CL). Since TL  is a necessary condition for an ideal BEC (as for any other ideal phase transition), this analysis is relevant in view of Section \ref{GRAV}, where we have studied a Bosonic gas in a reservoir, under the action of the weight. In contrast to what reported in the literature \cite{Huang,Goldstein,Lamb,deGroot,Gersch,Bagnato}, it has been shown that a rigorous BEC is excluded in any case, even if the approximate CL is adopted, to cure the discretness of the Airy spectrum, which survives TL. The study of the chemical potential below the ideal BEC temperature $T_f$ shows that the ground state population $N_0$ grows exponentially with decreasing temperature, just below $T_f$, but remains non-extensive, as one expects from an \textquoteleft approximate\textquoteright$\:$ BEC. A comparison with real experiments \cite{Condon, Rudolph} on a gas of $^{87}\mathrm{Rb}$, shows that replacing the confining harmonic trap with a rigid-wall cilinder (if that were possible) would make the increase of the condensate population a detrimentally smooth process. This shows the importance of the form of the gas-confining trap, in the experiments with ultra-cold atoms in the gravitational field.

\begin{appendices}
\numberwithin{equation}{section}

\section{Appendix A}
\label{Appendix A}
The classical phase space set $\Omega(E)$ (eqn \eqref{Omega}) is determined by the following conditions:
\begin{align}
&(x,\:y)\in\mathcal{A}\:;\quad0\le z \le L\:;\quad \epsilon_0\le \frac{p^2}{2m}+mgz\le E\nonumber\Rightarrow\\
\nonumber\\
&\Rightarrow
\begin{cases}
\sqrt{2m(\epsilon_0-mgz)}\le p \le \sqrt{2m(E-mgz)}&\:\text{for }mgz\le\epsilon_0\\
\label{CondOmega}\\
0\le p \le \sqrt{2m(E-mgz)}&\:\text{for }mgz\ge\epsilon_0
\end{cases}
\end{align}
\\
where $p=\sqrt{p^2_x+p^2_y+p^2_z}$. Due to the upper limit on $p\ge0$, it is necessary to separate the case $E\ge mgL$ from $E<mgL$, since in the former case $z$ must be integrated over the whole interval $[0,\:L]$, while in the latter the integration interval reduces to  $[0,\:E/(mg)]$. For compactness reasons, it is convenient to define the $E$-dependent lenght:

\begin{equation}
\label{elle(E)}
\ell(E):=
\begin{cases}
L\quad&\text{for }E\ge mgL\\
\\
E/(mg)\quad&\text{for }E\le mgL\:.
\end{cases}
\end{equation}
\\
On assuming, as obvious, $mgL>\epsilon_0$, from the conditions \eqref{CondOmega}, equation \eqref{N(E)} becomes: 

\begin{align}
&\mathcal{N}(E,\:L)=\frac{4\pi A}{\mathrm{h}^3}\times\nonumber\\
\nonumber\\
&\times\left[\int_0^{\epsilon_0/(mg)}\mathrm{d}z\int_{\sqrt{2m(\epsilon_0-mgz)}}^{\sqrt{2m(E-mgz)}}p^2\mathrm{d}p+\int_{\epsilon_0/(mg)}^{\ell(E)}\mathrm{d}z\int_0^{\sqrt{2m(E-mgz)}}p^2\mathrm{d}p\right]=\nonumber\\
\nonumber\\
&=\frac{4\pi (2m)^{3/2}A}{3\mathrm{h}^3}\Big[\int_0^{\epsilon_0/(mg)}\mathrm{d}z\left((E-mgz)^{3/2}-(\epsilon_0-mgz)^{3/2}\right)+\nonumber\\
\nonumber\\
&+\int_{\epsilon_0/(mg)}^{\ell(E)}\mathrm{d}z(E-mgz)^{3/2}\Big]=\nonumber\\
\nonumber\\
&=\frac{4\pi (2m)^{3/2}A}{3\mathrm{h}^3}\left[\int_{0}^{\ell(E)}\mathrm{d}z(E-mgz)^{3/2}-\int_0^{\epsilon_0/(mg)}\mathrm{d}z(\epsilon_0-mgz)^{3/2}\right]\:,\label{N(E,L)2}
\end{align}
\\
for $E\ge\epsilon_0$, while $\mathcal{N}(E,\:L)=0$ by definition for $E<\epsilon_0$. On noticing that the second integral in square brakets in eqn \eqref{N(E,L)2} is constant in $E$, and recalling eqn \eqref{elle(E)}, one gets:

\begin{align}
g_{acl}(E,\:L,\:g)&=\frac{\partial\mathcal{N}(E,\:L)}{\partial E}=\frac{2\pi (2m)^{3/2}A}{\mathrm{h}^3}\times\nonumber\\
\nonumber\\
&\times
\begin{cases}
\nonumber
\int_0^L(E-mgz)^{1/2}\mathrm{d}z&\quad\text{for }E\ge mgL\\
\\
\int_0^{E/(mg)}(E-mgz)^{1/2}\mathrm{d}z&\quad\text{for }\epsilon_0< E< mgL\\
\\
0&\quad\text{for }E< \epsilon_0\:.
\end{cases}
\end{align}
\\
From the preceding expression it is easy to derive eqn \eqref{G(E,L)}.

\section{Appendix B}
\label{Appendix B}

From eqns \eqref{total} and \eqref{Tf} it follows that, for $L\rightarrow\infty$:

\begin{align}
\label{AppB1}
&1=\frac{4\pi(8m)^{1/2}}{3g \mathrm{h}^3}\frac{A}{N}(\kappa T)^{5/2}\int_0^{\infty}\frac{(x+\beta\epsilon_0)^{3/2}}{\mathrm{e}^{x+\theta}-1}\mathrm{d}x\quad\Rightarrow\nonumber\\
\nonumber\\
\Rightarrow\quad&1=\overbrace{\left(\frac{1}{\Gamma(5/2)\zeta(5/2)}\right)}^{C}\left(\frac{T}{T_f}\right)^{5/2}\int_0^{\infty}\frac{(x+\beta\epsilon_0)^{3/2}}{\mathrm{e}^{x+\theta}-1}\mathrm{d}x\:,
\end{align}
\\
where $\theta:=\beta\Delta\mu$. In order to isolate the diverging part (for $\theta\rightarrow0$) in the r.h.s. member, we integrate by parts:

\begin{align}
1&=\overbrace{-C\left(\frac{T}{T_f}\right)^{5/2}(\beta\epsilon_0)^{3/2}\ln\left(1-\mathrm{e}^{-\theta}\right)}^{\text{diverging part}}-\nonumber\\
\label{div}\\
&\underbrace{-C\left(\frac{T}{T_f}\right)^{5/2}\frac{3}{2}\int_0^\infty(x+\beta\epsilon_0)^{1/2}\ln\left(1-\mathrm{e}^{-(x+\theta)}\right)\mathrm{d}x}_{\text{converging part}}\nonumber\:.
\end{align}
\\  
On expressing the converging part at the lowest order in $\beta\epsilon_0<<1$ and $\theta<<1$, equation \eqref{div} yields:

\begin{align}
1&=-Ct^{5/2}\Bigg[(r/t)^{3/2}\ln(\theta)+\nonumber\\
\label{div2}\\
&+\underbrace{\frac{3}{2}\int_0^\infty x^{1/2}\ln\left(1-\mathrm{e}^{-x}\right)\mathrm{d}x}_{-\Gamma(5/2)\zeta(5/2)}+\circ(r,\:\theta)\Bigg]\nonumber\:,
\end{align}
\\  
where $t:=T/T_f$, $r:=\epsilon_0/(\kappa T_f)$ and $\Gamma(\cdot)$ is Euler's function. The result expressed by the underbrace in eqn \eqref{div2} follows by integrating back by parts:

\begin{equation}
-\frac{3}{2}\int_0^\infty x^{1/2}\ln\left(1-\mathrm{e}^{-x}\right)\mathrm{d}x=\int_0^\infty \frac{x^{3/2}}{\mathrm{e}^{x}-1}\mathrm{d}x=\Gamma(5/2)\zeta(5/2)\:.
\end{equation}
\\
On recalling the definition of $T_f$ (eqn \eqref{Tf}), the preceding equation becomes:

\begin{equation}
-\ln \theta=\overbrace{\frac{1}{C}\left(\frac{\kappa T_f}{\epsilon_0}\right)^{3/2}}^{\mathcal{C}}\left[\frac{1-(T/T_f)^{5/2}}{T/T_f}\right]+\cdots\:\label{theta},
\end{equation}
\\
that will be used in the text (eqn \eqref{thetatext}).

\end{appendices}

\end{document}